\documentclass[final,3p,times]{elsarticle}

\usepackage{amssymb}
\usepackage{amsmath}

\journal{Journal of Subatomic Particles and Cosmology}

\begin{document}

\begin{frontmatter}



\title{The future fixed-target program at the CERN SPS}

\author[aaa]{Enrico Scomparin}
\affiliation[aaa]{organization={INFN Torino},
             addressline={V. P. Giuria 1},
             city={Torino},
             postcode={I-10125},
             country={Italy}}

\begin{abstract}
The CERN Super Proton Synchrotron (SPS) can currently accelerate heavy ions in the energy range from 13.5 up to 150 A GeV and deliver them to fixed-target experiments. It is the facility where Quark-Gluon Plasma (QGP) experimental studies began in 1986. After a first phase until 2000, with several fundamental discoveries made by a number of dedicated experiments, the physics program continued until today with the NA60 (dileptons, 2003-2004) and NA61 (hadronic observables, from 2009) experiments. As of today, a continuation of QGP studies, after the current shutdown of CERN accelerators, is foreseen, involving NA61 and a newly approved experiments, NA60+/DiCE. In this contribution, I will briefly describe the main achievements of the CERN SPS program, discuss the proposed new measurements and their impact on our knowledge of the QGP in the finite $\mu_{\rm B}$ region of the QCD phase diagram.
\end{abstract}

\end{frontmatter}



\section{Introduction}
\label{sec1}

The CERN SPS is ideally placed for the investigation of a wide region of the QCD phase diagram. For Pb-Pb collisions, energies in the center of mass $5<\sqrt{s_{\rm NN}}<17 $ GeV can be attained, allowing the exploration of a baryochemical potential range approximately corresponding to $220<\mu_{\rm B}<450$ MeV.
The first wave of experiments at this accelerator was concluded in 2000. Each experiment was typically concentrating on a selected range of topics, and the physics outcome of these experiments was summarized in a CERN press release, stating the presence of a ``compelling evidence for the existence of a new state of matter in which quarks, instead of being bound up into more complex particles such as protons and neutrons, are liberated to roam freely ''~\cite{CERNPressRelease}. Among the main discoveries, obtained by studying Pb-Pb collisions, one may underline the strangeness ``enhancement'', the ``anomalous'' suppression of the J/$\psi$, an excess in the real and virtual photon spectra, the thermal production of hadrons at chemical freeze-out and the radial flow.
After this pioneering phase, a second generation of experiments took place. The NA60 experiment (2003-2004) demonstrated the production of thermal dileptons in In-In collisions, with an average emission temperature exceeding the pseudo-critical temperature $T_{\rm c}$~\cite{Arnaldi:2008er}. Later on the NA61 experiment~\cite{NA61:2014lfx} performed, until today, detailed studies of hadron production, for a variety of nuclear collision energies, with the aim of investigating the properties of the onset of deconfinement and fireball, and to contribute to the search for the QCD critical point~\cite{NA61SHINE:2023epu}.

As of today (2026), the SPS can still deliver ion beams to fixed-target experiments, and will continue its operation at least until the end of the LHC running ($\sim 2041$). 
Therefore, the continuation of a heavy-ion related physics program at the SPS is technically possible. Among the reasons that may justify such a continuation one may cite:
\begin{itemize}
    \item the SPS energy domain was not thoroughly explored in the previous round of experiments, which where mainly carried out at top energy (158 GeV/nucleon for Pb-Pb collisions);
    \item the continuous improvement over $\sim$20 years of the detection techniques will place new experiments in a situation where for specific  measurements issues of limited accuracy may be overcome. A lesson from the past is illuminating in this respect: studies by NA60 in the electromagnetic probes domain (2003-2004) led to strong improvements with respect to previous decade experiments, thanks to the use of LHC-like Si detectors~\cite{NA60:2006ymb}.
\end{itemize}

This situation has triggered the upgraded proposal of the NA61/SHINE experiment~\cite{NA61SHINE:2025aey}, as well as of a completely new experiment, NA60+/DiCE~\cite{Ahdida:2932302}. The first one aims at investigating a rapid change in the features of hadron production between Be-Be (string-dominated) and Ar+Sc (collective/QGP-like), by studying collisions of light nuclei (O+O, Mg+Mg,B+B) at different energies, looking for the onset of deconfinement. The second has an ambitious program for the study of hard (open charm, charmonia) and electromagnetic probes of the QGP in Pb-Pb collisions, scanning the energy range accessible at the SPS. In the reminder of this contribution, I will analyze these projects and their potential influence on our knowledge of finite-$\mu_{\rm B}$ physics.

\section{Relevant physics topics and role of future SPS experiments}
The study of electromagnetic probes of the QGP features a specific sensitivity to the early phase of the collision, thanks to the absence of strong (re)interactions of the produced particles. In particular, the production of lepton pairs is a unique tool to determine the temperatures and lifetime of the strongly interacting medium created in the collisions. In addition, dileptons provide direct information about hadron spectral functions, in particular the $\rho$-meson, and its mixing with the chiral partner $a_{\rm 1}$, sensitive to chiral symmetry restoration which occurs in the vicinity of the transition from hadrons to a QGP. In the past, CERES~\cite{CERES:2006wcq} and NA60~\cite{NA60:2006ymb} have already given significant contributions in this domain. The extension of this physics program toward lower SPS energy presents an obvious interest, related, as an example, to the possibility of producing a ``caloric curve'' of the QGP, by studying the $\sqrt{s_{\rm NN}}$-dependence of the mass slope of dilepton spectra in the region between the $\phi$ and J/$\psi$ meson. This quantity corresponds to an average temperature of the system along its time evolution. Coupled with information from the forthcoming CBM experiment~\cite{Herrmann:2022jkv} at FAIR and higher-energy results from collider experiments~\cite{STAR:2024bpc}, one can investigate a possible anomalous behavior related to the onset of a first-order transition, that may become manifest in a flattening of the caloric curve. This feature might appear in the region close to the lowest energy accessible to the SPS, where a QCD critical point is expected.

Another relevant and little explored area in the SPS energy domain is heavy quark production. This class of ``hard probes'' of QGP was investigated in great detail by NA38/NA50/NA60 for what concerns J/$\psi$ and $\psi(2S)$ production at top SPS energy~\cite{Alessandro:2004ap}. Results for quarkonia at lower collision energy are currently unavailable for Pb-Pb collisions, and for open charm production no direct measurements exist 
The physics interest in this kind of studies is rather obvious: apart from the few results obtained at low energy,  extensive investigations at colliders have demonstrated the power of heavy-flavor (HF) particles to study key aspects of the medium: the diffusion of open HF particles enables direct access to a fundamental transport coefficient~\cite{ALICE:2021rxa}, while melting and regeneration of charmonium resonances give information on their in-medium binding potential~\cite{ALICE:2019lga}.

Concerning hadronic observables, studies at SPS energy on strangeness production, collective flow and fluctuations were already carried out to various extents.
In particular, fluctuation studies are relevant for the experimental search of the location of the QCD critical point (CP). Ratios of cumulants of net charge distribution were reported by NA61/SHINE, with hints to a possible non-monotonic behaviour~\cite{NA61SHINE:2025whi}. Moreover, studies by STAR in the frame of the RHIC-BESII program have shown, in the net proton cumulant ratio, a 5$\sigma$ deviation from reference calculations without CP at $\sqrt{s_{\rm NN}}$ = 19.6 GeV~\cite{STAR:2025zdq}. The energy range explored still includes a region (roughly between 4.5 and 7.5 GeV) where no data could be obtained. This region corresponds to the lower end of the SPS reach and upper end of FAIR domain and will be explored by these facilities in the future.

\begin{center}
\begin{figure}
    \centering
    \includegraphics[width=0.7\linewidth]{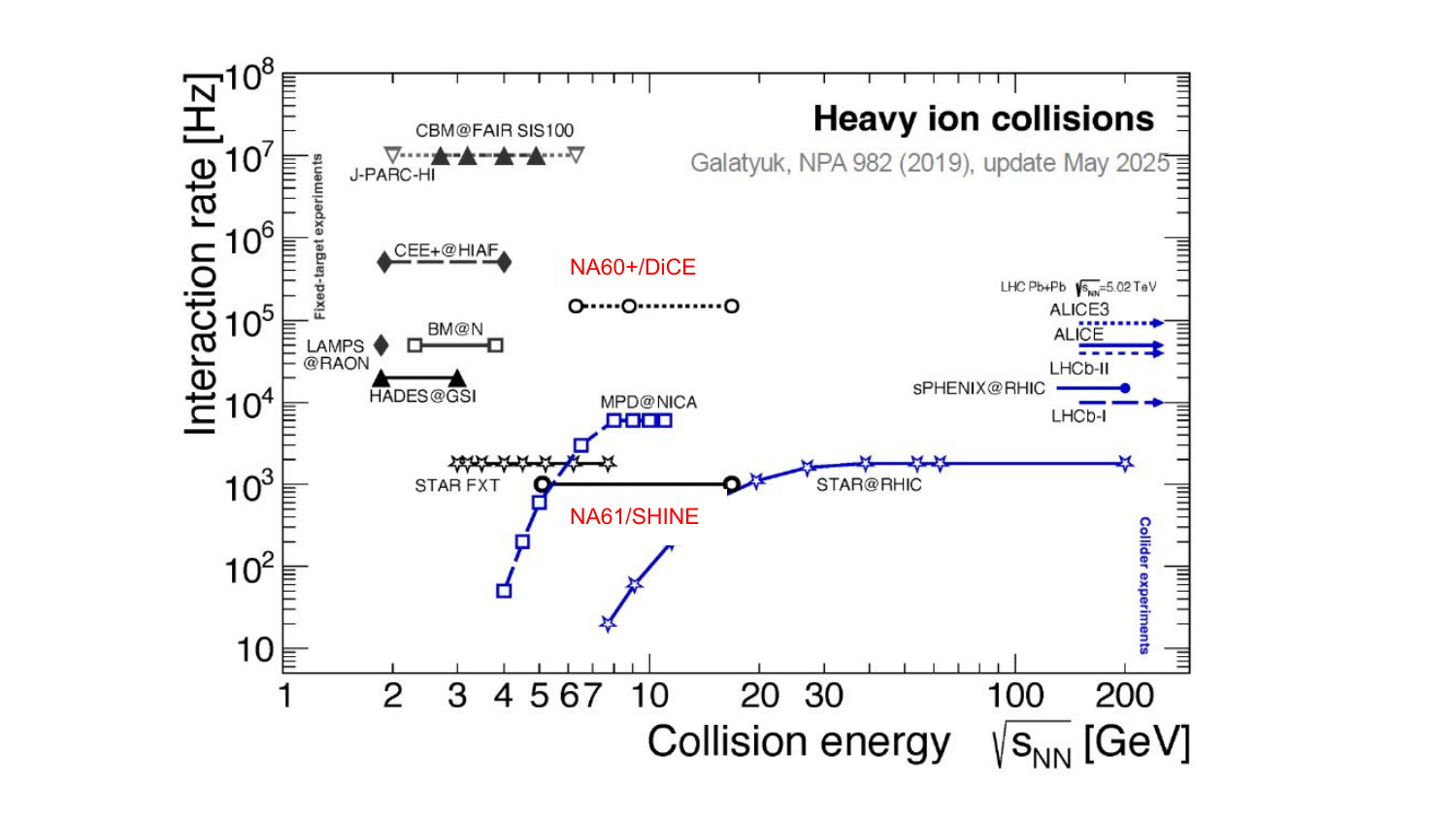}
    \caption{Center-of-mass energy coverage of present and future facilities for heavy-ion studies. The SPS experiments are marked in red.}
    \label{fig:facilities}
\end{figure}
\end{center}

Finally, the SPS represents a flexible and reliable facility for operation with heavy-ions. In addition to the wide energy range, also the instantaneous luminosity can vary over a large interval, reaching beam intensity larger than 10$^6$ Pb ions/s. This brings a significant opportunity for accurate measurements of low-cross section processes, as it is the case for hard and electromagnetic probes of the QGP. Indeed, as can be seen in Fig.~\ref{fig:facilities}~\cite{Galatyuk:2019lcf}, where the possible interaction rates at various facilities are shown, the only potential competing facility in the same energy range (NICA~\cite{Kolesnikov:2020qfw}), can reach at most an interaction rate smaller by more than one order of magnitude.

\section{Prospects for NA61/SHINE}
Started in 2009, the physics program of the NA61/SHINE experiment is multi-faceted, including the exploration of strongly interacting matter, the investigation of neutrino oscillations, and the interpretation of cosmic-ray air showers, via the measurement of hadron production in the SPS energy range. Among its highlights, the experiment has confirmed and mapped the ``horn'' structure, a non-monotonic behavior in the \(K^{+}/\pi ^{+}\) multiplicity ratio as a function of collision energy. This result was interpreted as a key signature of the transition threshold toward a QGP~\cite{NA61SHINE:2023epu}.

In 2023, NA61/SHINE submitted a document to the SPSC for a light-ion program after the Long Shutdown 3~\cite{NA61SHINE:2025aey}. The proposal includes measurements with oxygen, magnesium, and boron beams at three momenta: 13, 30, and 150AGeV/c. These measurements aim at exploring the system-size and energy dependence of the transition from hadronic to partonic matter, by identifying the onset of QGP production at the top SPS energy, studying the interplay of hadronic and partonic mechanisms at intermediate energy (30AGeV/c), and investigating resonance-dominated dynamics at the lowest energy (13AGeV/c). On the hardware side, there are plans for the replacement of the first large-volume TPC with a fast tracking system allowing operation at up to $\sim 10$ kHz, based on 
Si-pixel detectors close to the target and larger-area gas detectors positioned downstream.
One of the important goals is to perform, in addition to the light-ion program, first direct measurements of charm-hadron production in central Pb-Pb collisions, including studies of charm–anticharm correlations. Multi-strangeness production studies will also be part of this upgraded program. 

\section{The new NA60+/DiCE experiment}

\begin{center}
\begin{figure}
    \centering
    \includegraphics[width=0.7\linewidth]{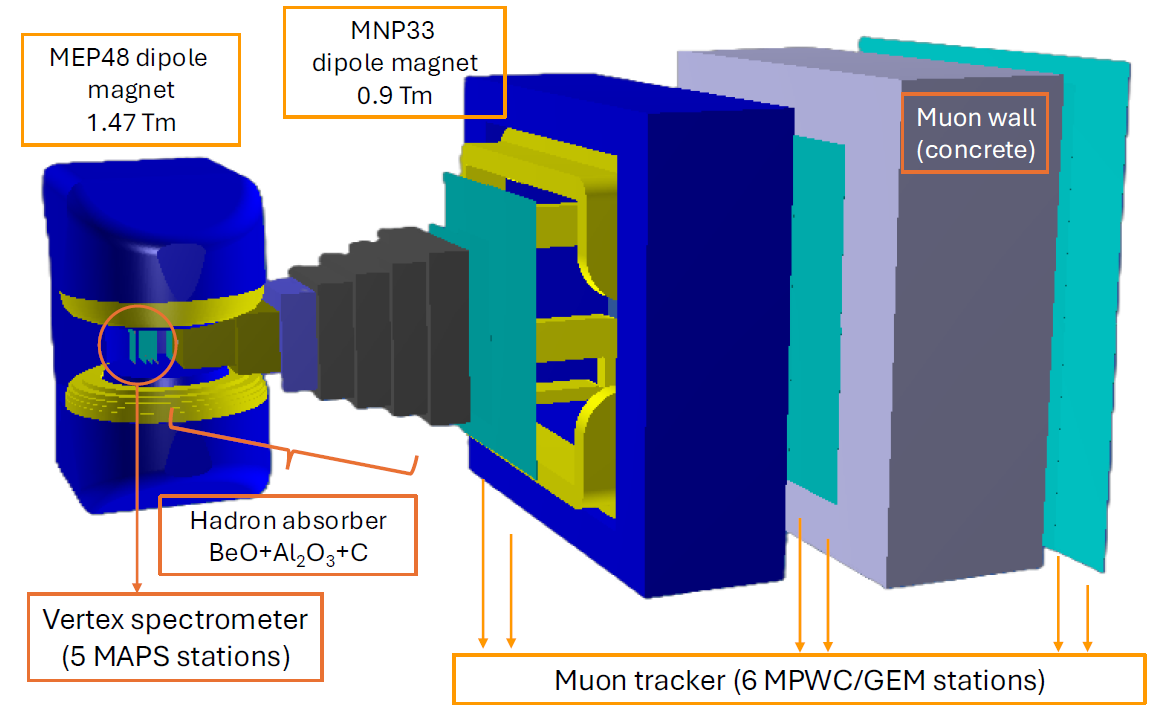}
    \caption{The experimental set-up of the NA60+/DiCE experiment. Details are given in the text.}
    \label{fig:DiCE_setup}
\end{figure}
\end{center}

Studies for a new high-precision experiment devoted to the study of the hard and electromagnetic probes began in the previous decade. In recent years, a proposal~\cite{Ahdida:2932302} was submitted to the SPS experimental committee (SPSC) and finally the experiment was approved by CERN in June 2026. The experimental set-up, shown in Fig.~\ref{fig:DiCE_setup}, is inspired by the former NA60 detector. It includes a vertex spectrometer and a muon spectrometer. The two detectors are separated by a thick hadron absorber and each of them features a dipole magnet. For the vertex spectrometer this is the MEP48 magnet, currently stored at CERN, which generates a $\sim 1.5$ T field over a 40 cm wide gap. The muon spectrometer will be equipped with the MNP33 magnet, currently in use (until 2026) by the CERN NA62 experiment. This is a wide gap ($\sim 240\times240$ cm$^2$) dipole, with an integrated strength of $0.9$ T$\cdot$m, its field extending much beyond the 1.3 m dimension of the yoke.
The vertex spectrometer is based on Monolithic Active Pixel Sensors (MAPS), organized in five stations of increasing dimension from the upstream to the downstream ones with respect to the beam direction. Each station is made of four large sensors, organized in two planes with two sensors each, staggered in such a way to limit the occurrence of dead zones. Each large sensor is produced using a stitching technique that allows the replication of smaller sensors (Repeated Sensor Units, RSU) and the transfer of data toward the periphery of the large sensor. The design was originally developed in the frame of the ALICE ITS3 project~\cite{The:2890181}. Rows of seven RSU are produced and each large sensor is formed by 3 to 7 of such rows. The larger sensor has a dimension of $\sim 13.5\times 13.5$ cm$^2$.

The muon spectrometer is made by six MWPC tracking stations, of increasing transverse dimensions, with the MNP33 magnet being positioned between the second and third station. The two last tracking stations are preceded by a second absorber, made of concrete and 160 cm thick, that stops punch-through hadrons from the main absorber. A low- and a high-energy configuration of the muon spectrometer are foreseen, the latter featuring a thicker hadron absorber and a downstream shift of the detectors and magnet, to keep a constant rapidity acceptance and ensure an efficient removal of the larger hadronic multiplicity. The relatively low rate expected from FLUKA simulations in the muon spectrometer stations ($< 2$ kHz/cm$^2$) makes the choice of MWPCs multi-wire proportional chambers with strip readout attractive, as highly reliable, large area, nearly 100\% efficient detectors can be built at low cost. Depending on further assessments, GEM or $\mu$RWell detectors will be considered for the most upstream station, where hadronic multiplicity is larger.

By matching tracks in the muon spectrometer with corresponding tracks in the vertex spectrometer, in position and momentum space, a high resolution on the muon measurement can be obtained, as well as a precise determination of the origin of the muons with respect to the position of the primary vertex. More in detail, a resolution of 11 (33) MeV on the mass of the $\omega$ (J/$\psi$) meson, detected via its dimuon decay, is expected. As a comparison, the former NA60 experiment reached a 23 MeV resolution at the $\omega$. The expected resolution on the impact parameter of the tracks is $\sim 20 \mu$m at $p_{\rm T}=1$ GeV/$c$, showing that the identification of decay products of charmed hadrons is possible.

\begin{figure}[ht!]
\begin{center}
\includegraphics[width=0.4\textwidth]{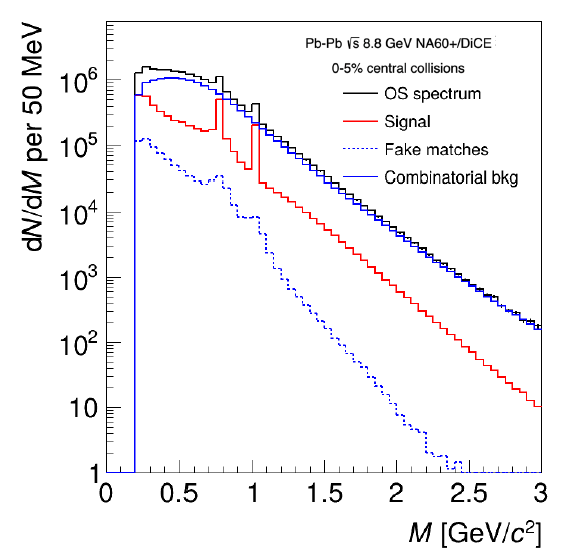}
\includegraphics[width=0.4\textwidth]{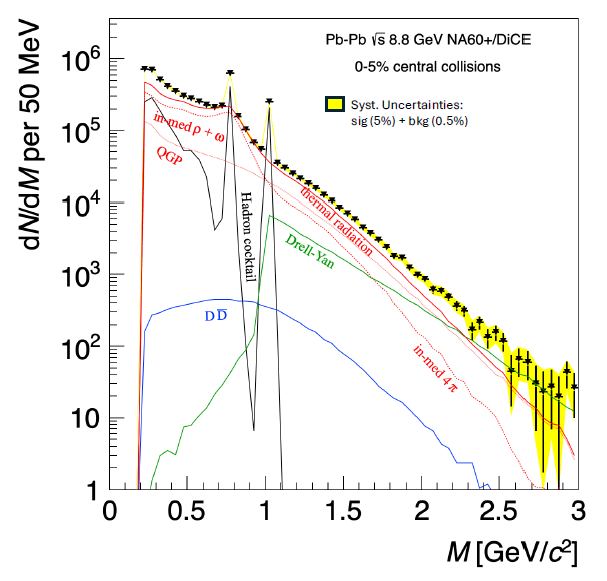}
\caption{Expected dimuon sample in the 5\% most central Pb--Pb collisions at $\sqrt{s_{\rm NN}} = 8.8$ GeV before (left) and after (right) subtraction of combinatorial and fake match background.
The various contributions are shown (see text for details). }
\label{fig:DileptonsDiCE}
\end{center}
\end{figure}

Detailed physics performance studies have then been carried out. Concerning electromagnetic probes, in Fig.~\ref{fig:DileptonsDiCE} (left) we show the total reconstructed mass spectrum (black) corresponding to a data sample collected in one month of data taking at $\sqrt{s_{\text{NN}}} = 8.8$ GeV, where the expected total integrated luminosity is $0.6 \times 10^{12}$ ions on target. After the subtraction of the combinatorial background and a small contribution of fake matches, one is left with the signal spectrum shown in Fig.~\ref{fig:DileptonsDiCE} (right). All the expected signal components are shown. For $m_{\rm \mu\mu} < 1$ GeV/c$^2$, the thermal radiation yield is dominated by the in-medium $\rho$ decay.  The thermal spectrum is measurable up to 2.5--3 GeV/c$^2$. The open-charm contribution to the dimuon spectrum  becomes totally negligible at low $\sqrt{s_{\text{NN}}}$. The Drell--Yan yield will be measured in dedicated p-A runs. After acceptance correction, a fit of the spectrum with $dN/dM \propto M^{3/2} \exp(-M/T_{\text{slope}})$ in the interval $1.5<m_{\rm \mu\mu} <2.5$ GeV/c$^2$ allows the extraction of the average temperature, with a projected total uncertainty of $\sim 4$\%. This demonstrates the possibility of detecting structures in the $\sqrt{s_{\rm NN}}$-dependence of $T_{\text{slope}}$ possibly related to the order of the phase transition.

In addition to thermal production, the dilepton invariant mass spectrum can also be sensitive to a signal of chiral symmetry restoration. A change in the degree of mixing of the chiral partners $\rho$ and $a_1$ can be shown to lead to an increase of the dilepton yield, concentrated in the region $0.9<m_{\mu\mu}<1.4$ GeV/$c^2$, where in absence of such an effect a dip of the vector spectral function is present~\cite{Hohler:2013eba}. The expected total uncertainties show a very good sensitivity to an increase of the yield due to chiral mixing of ${\sim}20\text{--}30\%$.

\begin{figure}[ht!]
\begin{center}
\includegraphics[width=0.4\textwidth]{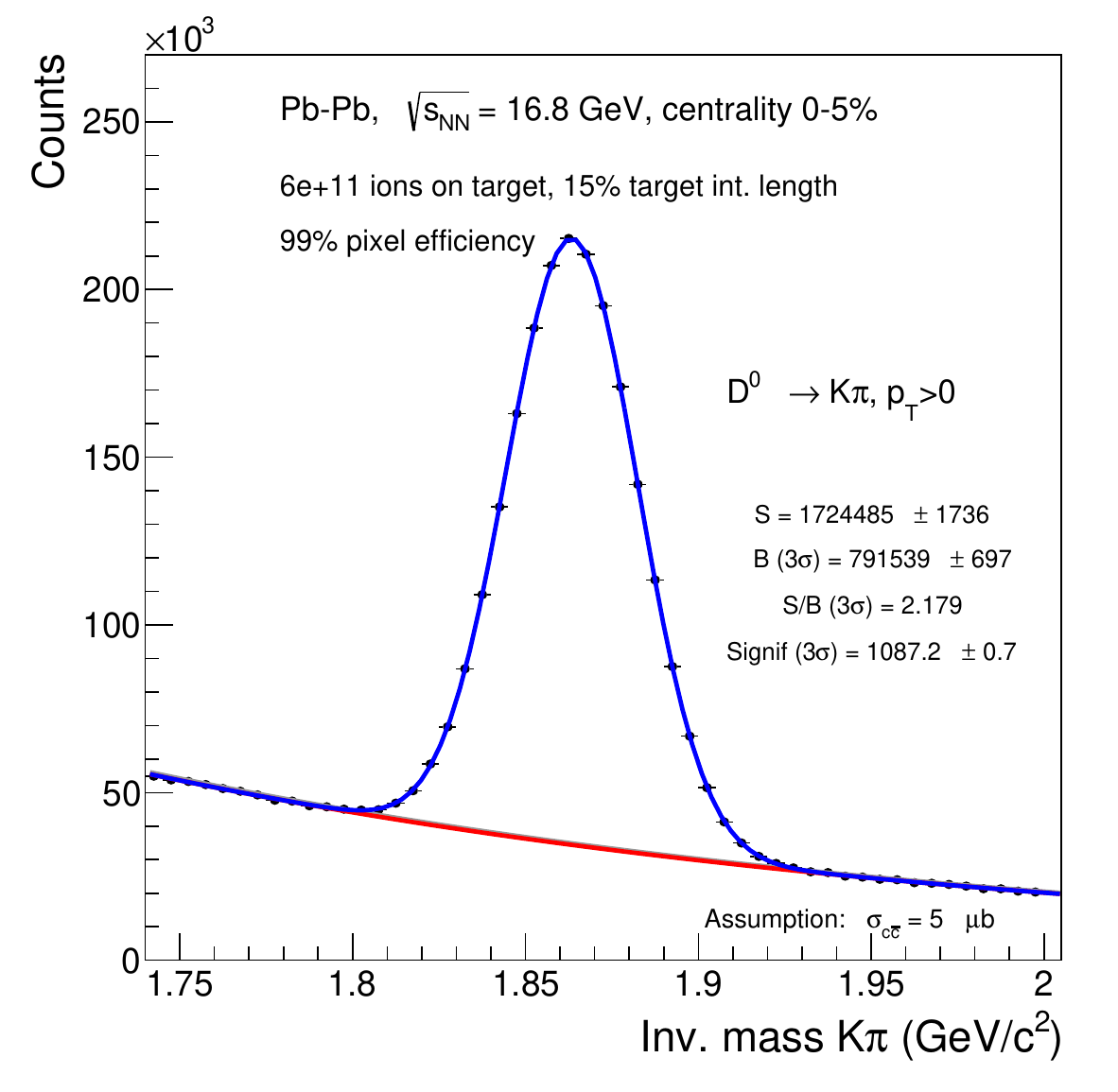}
\includegraphics[width=0.42\textwidth]{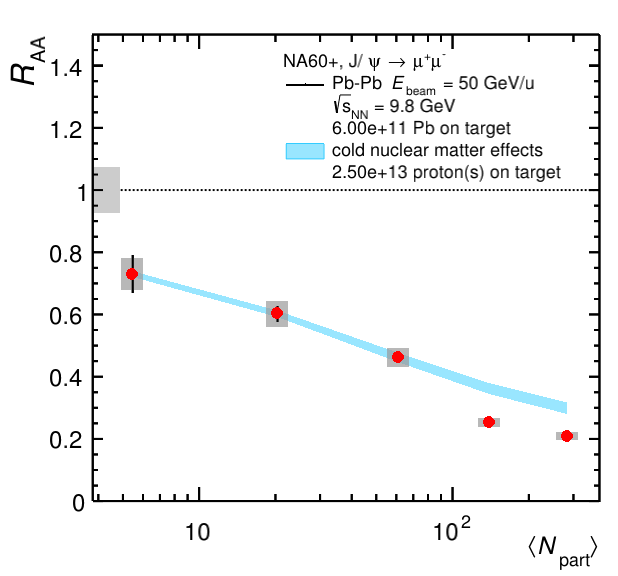}
\caption{Expected invariant-mass distribution of D$^0$ candidates in $5\cdot 10^9$ central Pb-Pb collisions at beam energy of 150 A GeV (left); simulated performance for the J/$\psi$ $R_{\rm AA}$  in Pb-Pb collisions at $E_{\rm lab}=50$A GeV, as a function of centrality. A 30\% suppression, in addition to cold-nuclear matter effects, is assumed in the two more central bins (right). }
\label{fig:D0}
\end{center}
\end{figure}

Moving to heavy-quark production, open charm hadrons can be reconstructed with the NA60+/DiCE experiment through their decays into two or three charged hadrons.  These hadrons are detected by reconstructing their tracks in the vertex telescope. For this observable, the information of the muon spectrometer is obviously not used. The large combinatorial background is reduced through geometrical selections based on the displaced decay-vertex topology, exploiting the fact that the mean proper decay lengths $c\tau$ of open charm hadrons range from 60 to 310~$\mu$m depending on the hadron species. Even without particle-identification capabilities, a clear identification of D$^0$, D$_{\rm s}$ and $\Lambda_{\rm c}$ can be obtained. In Fig.~\ref{fig:D0} (left) the expected invariant mass distribution of D$^0$ candidates in the 5\% most central Pb-Pb collisions at $E_{\rm lab}=150 A$ GeV is shown. This plot corresponds to a sample of $6 \times 10^{11}$ ions on target, i.e., one month of data taking. The background was simulated starting from a parameterization of NA49 data.

In Fig.~\ref{fig:D0} (right) the simulated performance for a measurement of the J/$\psi$ $R_{\rm AA}$ in Pb-Pb collisions at $E_{\rm lab}=50$ GeV is shown. The integrates luminosity corresponds to one month of data taking. Cold nuclear matter effects are evaluated assuming a break-up cross section of 7.6 mb, as measured by NA60 in p-A collisions at 158 GeV~\cite{NA60:2010wey}, and a further 30\% suppression, intended to simulate QGP effects, is included for the two more central Pb-Pb points. 
 
\section{Conclusions}
The CERN SPS is a powerful and versatile facility in the energy range $5<\sqrt{s_{\rm NN}}<17$ GeV, that offers significant possibilities for QGP studies at finite $\mu_{\rm B}$. In particular, little is known below top SPS energy on hard and electromagnetic probes. Their penetrating nature makes them ideal for the characterization of a system that spends significant time in the region around deconfinement and chiral transition. A brand-new experiment, NA60+/DiCE, was recently approved by CERN, with a physics program spanning the next decade, after the completion of Long Shutdown 3. Also, NA61/SHINE is proposing a continuation of its systematic measurements of hadronic observables, mainly with light-ion collisions. A support of this program, both in terms of new 
theoretical ideas and developments, and contribution of our community to the experimental efforts, will contribute to guarantee a very successful physics outcome.

\bibliographystyle{elsarticle-num}
\bibliography{sn-bibliography}



\end{document}